\documentclass[preprint,floatfix]{revtex4}
\usepackage[latin9]{inputenc}
\usepackage{amsmath}
\usepackage{amssymb}
\usepackage{graphicx}
\usepackage{esint}

\makeatletter

\providecommand{\tabularnewline}{\\}

\@ifundefined{textcolor}{}
{%
 \definecolor{BLACK}{gray}{0}
 \definecolor{WHITE}{gray}{1}
 \definecolor{RED}{rgb}{1,0,0}
 \definecolor{GREEN}{rgb}{0,1,0}
 \definecolor{BLUE}{rgb}{0,0,1}
 \definecolor{CYAN}{cmyk}{1,0,0,0}
 \definecolor{MAGENTA}{cmyk}{0,1,0,0}
 \definecolor{YELLOW}{cmyk}{0,0,1,0}
}

\usepackage{epsfig}

\makeatother

\begin{document}

\title{Analytical approach to the $D3$-brane gravity dual for $3d$ Yang-Mills
theory}

\author{Hilmar Forkel$^{1,2}$ }

\affiliation{$^{1}$Departamento de F�sica, Instituto Tecnol�gico de Aeron�utica,
12228-900 S\~{a}o Jos\'{e} dos Campos, S\~{a}o Paulo, Brazil}

\affiliation{$^{2}$Institut f�r Theoretische Physik, Universit�t Heidelberg,
D-69120 Heidelberg, Germany}
\begin{abstract}
The complexity of ``top-down'' string-dual candidates for strongly-coupled
Yang-Mills theories and in particular for QCD almost always prohibits
their exact analytical or even comprehensive numerical treatment.
This impedes both a thorough quantitative analysis and the development
of more realistic gravity duals. To mitigate these impediments, we
devise an analytical approach to top-down duals on the basis of controlled,
uniformly converging high-accuracy approximations for the normalizable
string modes corresponding to gauge-theory states. We demonstrate
the potential of this approach in Witten's dual for $3d$ Yang-Mills
theory, i.e. in the near-horizon limit of non-extremal $D\text{3}$-branes,
compactified on $S^{1}$. We obtain accurate analytical approximations
to the bulk modes which satisfy the boundary conditions exactly. On
their basis, analytical results for masses, sizes, pole residues and
correlation functions of glueball excitations are derived by spectral
methods. These approximations can be systematically improved and rather
universally adapted to more complex gravity duals.
\end{abstract}
\maketitle

\section{Introduction}

Since their inception nearly two decades ago, gauge/string dualities
keep generating profound insights into strongly-coupled gauge theories
and have led to new perspectives and groundbreaking results over a
broad range of physics subfields \cite{revs}. 

Among the pioneering and most intensely studied application areas
for these holographic dualities is QCD, the theory of the strong interactions,
in its infrared regime. Hence, various increasingly sophisticated
and realistic string-theory-based (i.e. ``top-down'') dual candidates
for strongly-coupled large-$N$ Yang-Mills theories and QCD were constructed.
In order to be at least halfway accessible by contemporary solution
methods, most of them use the gravity approximation for the string
dynamics %
\footnote{ or on highly rotationally excited strings%
}. This requires weak bulk spacetime curvatures and entails several
well-understood limitations \cite{revs} e.g. in the description of
asymptotic freedom or linear Regge trajectories. 

Even within these restrictions, however, the technical (and sometimes
conceptual) complexities encountered when attempting to calculate
gauge-theory quantities beyond approximate mass spectra can be quite
formidable. Part of these problems arise from the absence of analytical
solutions for even the simplest bulk quantities in practically all
top-down duals. Moreover, numerical approaches provide no fully satisfactory
alternative. The numerical evaluation of more complex amplitudes like
correlators is e.g. typically hampered by renormalization issues and
often requires specifically designed, non-universal methods. In addition,
numerical results reveal less of the underlying mathematical structure
whose patterns and systematics often provide useful insights and intuitive
guidance.

To improve this situation, we intend to develop and adapt more sophisticated
analytical approximation methods to deal with top-down gravity duals.
They should be controlled (i.e. systematically improveable) and as
transparent and universally applicable as possible. In particular,
they should provide accurate analytical expressions for a complete
set of dual quantities from which analytical results for all desired
gauge-theory observables can be derived. In this letter we propose
the whole tower of (normalizable) dual string modes, corresponding
to the ``hadrons'' of interest, as a promising such set.

In the following, we demonstrate the feasibility of our approach in
Witten's top-down gravity dual for three-dimensional Yang-Mills theory
(YM$_{3}$) \cite{wit98,bro00,csa99} which emerges from $N$ parallel,
non-extremal $D3$ branes of type IIB string theory in the near-horizon
limit, with one of its flat dimensions compactified on a circle. This
dual model suggests itself for our purposes because it is technically
simple and reproduces important properties of the gluonic sector of
QCD (including confinement).

\section{Gravity dual for $3d$ Yang-Mills theory}

In Ref. \cite{wit98}, Witten proposed a string dual for $SU\left(N\right)$
Yang-Mills theory in $d=2+1$ dimensions at large $N$ and for large
't Hooft coupling $\lambda=g_{YM}^{2}N$. This dual is based on the
AdS/CFT correspondence between $\mathcal{N}=4$ superconformal Yang-Mills
theory in $d=3+1$ dimensions and a stack of $N$ coincident $D3$-brane
solutions of type IIB string theory in 10 dimensions (with R-R charge\textbf{
$N$}), or for small spacetime curvature by IIB supergravity \cite{mal97}.
Conformal symmetry and supersymmetry are broken by generalizing the
correspondence to non-extremal $D3$ black-brane solutions with\textbf{
}Hawking temperature $T_{H}$. 

One then decouples the near-horizon or ``throat'' region from the
10d Minkowski gravity (staying outside the horizon) and analytically
continues to Euclidean time $\tau=-it$. This requires to periodically
identify $\tau+\beta_{H}\simeq\tau$ on the circle $S^{1}\left(R_{\tau}\right)$
and to remove a cone singularity in the $\left(\tau,r\right)$ subspace
by fixing the Hawking temperature to $T_{H}=\left(\pi R\right)^{-1}$
with the horizon at $r_{0}\simeq\pi T_{H}R^{2}=R$. In the Fefferman-Graham
coordinate $z=R^{2}/r$, the metric then becomes 
\begin{equation}
ds^{2}=g_{MN}dx^{M}dx^{N}=\frac{R^{2}}{z^{2}}\left[\left(1-\frac{z^{4}}{R^{4}}\right)d\tau^{2}+\sum_{i=1,2,3}dx_{i}^{2}+\left(1-\frac{z^{4}}{R^{4}}\right)^{-1}dz^{2}\right]+R^{2}d\Omega_{5}^{2}\label{eq:ads-bb}
\end{equation}
where $x_{3}$ is reinterpreted as the Euclidean time. (The associated
dilaton is constant.)

On the gauge-theory side the above procedure corresponds to compactifying
the Euclidean $\mathcal{N}=4$ $SU\left(N\right)$ Super-Yang-Mills
theory (i.e. the low-energy theory of the $N$ coinciding $D3$ branes)
on $\mathbb{R}^{3}\times S_{\tau}^{1}$, with anti-periodic, supersymmetry-breaking
boundary conditions for the fermions around $S_{\tau}^{1}$. Hence
the adjoint fermions acquire tree-level masses $\sim1/R_{\tau}$ (thus
breaking conformal symmetry) and render the scalar matter massive
at the one-loop level. For sufficiently small $R_{\tau}$ the adoint
matter decouples and only the gauge fields on $\mathbb{R}^{3}$ remain
massless (protected by the gauge symmetry), with their dynamics approaching
YM$_{3}$ (with an UV cutoff $\sim1/R_{\tau}$). 

The background geometry (\ref{eq:ads-bb}) is dual to the gauge-theory
vacuum. Its scalar excitations, the $0^{++}$ ``glueballs'', are
associated with the (gauge-invariant) operator 
\begin{equation}
\mathcal{O}{}_{0^{++}}\left(x\right)=tr\left\{ F_{\mu\nu}F^{\mu\nu}\right\} \label{eq:GbOp}
\end{equation}
which in turn is dual to a bulk field with identical quantum numbers.
The corresponding deformation $S\rightarrow S+\int d^{d}x\varphi^{\left(s\right)}\left(x\right)\mathcal{O}{}_{0^{++}}\left(x\right)$
of the gauge-theory action changes the Yang-Mills coupling $g_{YM}^{2}=g_{s}=\exp\varphi^{\left(s\right)}$
and thus the (boundary) value $\varphi^{\left(s\right)}$ of the dilaton
field $\varphi$. This indicates that $\varphi$ is the bulk field
dual to $\mathcal{O}{}_{0^{++}}$. $ $

The GKPW relation \cite{gub98} then identifies the generating functional
of the glueball correlation functions as 
\begin{equation}
Z\left[\varphi^{\left(s\right)}\right]=\left\langle e^{\int d^{4}x\varphi^{\left(s\right)}\left(x\right)\mathcal{O}{}_{0^{++}}\left(x\right)}\right\rangle \overset{N\gg\lambda\gg1}{\longrightarrow}e^{-S^{\left(onsh\right)}\left[\varphi^{\left(s\right)}\left(x\right)=\varphi\left(z=0,x\right)\right]}\label{eq:gfun}
\end{equation}
with the string partition function in the classical limit. The latter
is determined by the on-shell action $S^{\left(onsh\right)}$, i.e.
by the classical dilaton action (after dimensional reduction on $S^{5}$,
i.e. for $\varphi$ restricted to SO$\left(6\right)$ singlets) 
\begin{align}
S_{0,m}\left[\varphi\right] & =\frac{1}{2\kappa^{2}}\int d^{5}x\sqrt{\left|g\right|}\left[g^{MN}\partial_{M}\varphi\partial_{N}\varphi+m_{5}^{2}\varphi^{2}+O\left(\alpha^{'n}\right)\right]\label{eq:DilatonAction}
\end{align}
($\kappa^{2}/2$ is the five-dimensional Newton constant) in the dual
background geometry (\ref{eq:ads-bb}) evaluated at the extrema of
Eq. (\ref{eq:DilatonAction}). The latter are the solutions of the
Laplace-Beltrami equation 
\begin{equation}
\left(g^{\tau\tau}\partial_{\tau}^{2}+\Delta_{z}+g^{ij}\partial_{i}\partial_{j}-m_{5}^{2}\right)\varphi\left(x,\tau,z\right)=\int d^{3}ke^{iqx}\left(\Delta_{z}-a^{-2}q^{2}-m_{5}^{2}\right)\varphi\left(q;z\right)=0\label{eq:LBeq}
\end{equation}
after dimensional reduction on $S^{1}\left(R_{\tau}\right)$, i.e.
for $\tau$-independent $\varphi$. Here $\varphi\left(q;z\right)$
is the $3d$ Fourier transform of the flat boundary coordinates of
the dilaton field and
\begin{align}
\Delta_{z} & \equiv\frac{1}{\sqrt{g}}\partial_{z}\sqrt{g}g^{zz}\partial_{z}=\frac{z^{5}}{R^{5}}\partial_{z}\left(\frac{R^{3}}{z^{3}}-\frac{z}{R}\right)\partial_{z}
\end{align}
is the radial Laplacian. A Frobenius analysis shows that the normalizable
(non-normalizable) solutions of Eq. (\ref{eq:LBeq}) behave as $z^{d}$
($z^{\Delta-d}$) near the boundary. For the (classical) conformal
dimension $\Delta=4$ of the interpolator (\ref{eq:GbOp}) the AdS/CFT
dictionary implies $m_{5}^{2}R^{2}=\Delta\left(\Delta-d\right)\overset{\Delta=d=4}{=}0$,
as expected for massless dilaton modes.

As motivated in the introduction, we intend to construct approximate
solutions of the radial dilaton equation 
\begin{equation}
\left[a^{2}\left(z\right)\Delta_{z}-q^{2}\right]\varphi\left(q;z\right)=0\label{eq:radeq0}
\end{equation}
which are normalizable and regular at the horizon. Normalizability
selects the subleading solutions which obey the UV boundary condition
$\varphi\left(q;z\right)\overset{z\rightarrow0}{\longrightarrow}z^{4}$.
These solutions have discrete eigenvalues $q_{n}^{2}=-m_{n}^{2}$
for $n=1,2,...$ where $m_{n}$ is the mass of the $n$-th radial
glueball excitation \cite{wit98}. More explicitly, the $n$-th eigensolution
$\psi_{n}\left(z\right):=\varphi\left(q_{n};z\right)$ solves the
radial equation
\begin{equation}
\left[\partial_{z}^{2}+t_{1}\left(z\right)\partial_{z}+m_{n}^{2}t_{2}\left(z\right)\right]\psi_{n}\left(z\right)=0\label{eq:radeq}
\end{equation}
(a 2nd-order Fuchsian differential equation) whose coefficient functions
\begin{equation}
t_{1}\left(z\right)=-\frac{z^{4}+3R^{4}}{z\left(R^{4}-z^{4}\right)},\qquad t_{2}\left(z\right)=\frac{R^{4}}{R^{4}-z^{4}}
\end{equation}
have four regular singular points at $z=0,\pm R,\infty$. Hence no
general, analytical solutions for Eq. (\ref{eq:radeq}) are known.

\section{Uniform analytical approximations to the string modes}

In this section, we will derive controlled, uniformly convergent analytical
approximations to all eigenmodes $\psi_{n}$ of Eq. (\ref{eq:radeq})
under the stated boundary conditions. Since we are dealing with a
Sturm-Liouville problem which is not of the standard Schr�dinger type
(see below), we will rely on a generalization of the WKB method (GWKB)
\footnote{We note that next-to-leading order WKB estimates of eigenvalue (i.e.
mass) spectra are known for several gravity duals, including the one
we are considering here (cf. e.g. Refs. \cite{min99,wkb-spectra,bro00,bro200}).
We propose N$^{2}$LO approximations to the\emph{ eigenmodes}, however,
which contain much more information (see below).%
} and just sketch the main steps here, relegating more details to a
subsequent, more comprehensive publication \cite{for16}. 

To impose the boundary conditions at the singular points $z=0,R$,
it is convenient to transform to the new radial coordinate
\begin{equation}
y\left(z\right)=\ln\left(\frac{R^{2}}{z^{2}}-1\right)\label{eq:y(z)}
\end{equation}
which maps the horizon to $y=-\infty$ and the AdS boundary to $y\rightarrow+\infty$.
Equation (\ref{eq:radeq}) is thereby converted into the quasi-Schr�dinger
form
\begin{equation}
\left[-\partial_{y}^{2}+\frac{1}{\epsilon^{2}}V_{1}\left(y\right)+V_{2}\left(y\right)\right]u\left(y\right)=0.\label{eq:radeq2}
\end{equation}
with the potentials 
\begin{align}
V_{1}\left(y\right) & =-\frac{e^{y}}{\left(1+e^{y}\right)\left(2+e^{y}\right)}\label{eq:v1}
\end{align}
and
\begin{equation}
V_{2}\left(y\right)=\frac{1}{2}\frac{\left(3+4e^{y}\right)e^{y}}{\left(1+e^{y}\right)\left(2+e^{y}\right)}-\frac{1}{4}\frac{\left(3+2e^{y}\right)^{2}e^{2y}}{\left(1+e^{y}\right)^{2}\left(2+e^{y}\right)^{2}}\label{eq:v2}
\end{equation}
where the eigenvalues appear in the combination $\epsilon^{-2}=\left(m^{2}R^{2}\right)/4$. 

A global and uniform (asymptotic) expansion of the solutions of Eq.
(\ref{eq:radeq2}) can be organized in the familiar form
\begin{equation}
u\left(y\right)\sim\exp\left(\frac{1}{\epsilon}\sum_{k=0}^{\infty}S_{k}\left(y\right)\epsilon^{k}\right),\qquad\epsilon\rightarrow0+\label{eq:GWKBansatz}
\end{equation}
whose $\epsilon$ dependence implements a standard version of ``distinguished
balance'' \cite{ben78}. The leading-order solution is thus of $O\left(\epsilon^{0}\right)$.
Equation (\ref{eq:radeq2}) implies that the potentials $V_{1,2}$
contribute at different orders to the asymptotic expansions. This
requires a non-standard two-scale power counting analysis and the
corresponding GWKB matching procedure around the turning points of
Eqs. (\ref{eq:v1}) and (\ref{eq:v2}). 

For approximations that are sufficiently accurate to satisfy both
the\textbf{ }regularity (\textbf{$u\left(-\infty\right)=c$)} and
normalizability ($u\left(+\infty\right)=\lim_{y\rightarrow\infty}ce^{-y}$)
boundary conditions exactly, one has to push the expansion (\ref{eq:GWKBansatz})
to $O\left(\epsilon^{2}\right)$, i.e. to next to next to leading
order (N$^{2}$LO). A uniform approximation further requires to match
$u\left(y\right)$ near both boundaries to exact solutions of approximate
radial field equations, expanded to the appropriate orders of $e^{y}$
and $\epsilon$. (This becomes necessary since to N$^{2}$LO the ansatz
(\ref{eq:GWKBansatz}) diverges at both boundaries\textbf{ $y\rightarrow\pm\infty$}.)
The resulting two matching relations (or, equivalently, the two boundary
conditions) can be simultaneously satisfied only for the discrete
spectrum 
\begin{equation}
\epsilon_{n}=\frac{2}{m_{n}R}=\frac{2\,_{2}F_{1}\left(\frac{1}{4},\frac{1}{2},\frac{5}{4},1\right)}{\left(n+\frac{1}{2}\right)\pi}\overset{n\gg1}{\longrightarrow}\frac{2^{-1/2}\sqrt{\pi}}{\Gamma\left(\frac{3}{4}\right)^{2}\sqrt{n\left(n+1\right)}},\qquad n=1,2,...\label{eq:EVcond}
\end{equation}
of eigenvalues (where $\,_{2}F_{1}$ is Gauss' hypergeometric function
\cite{abr72}). Since Eq. (\ref{eq:EVcond}) does not receive N$^{2}$LO
corrections, its large-$n$ version can alternatively be obtained
from an NLO Bohr-Sommerfeld type integral between the two turning
points of $V$ \cite{min99}. 

After transforming the \textbf{$u_{n}^{\left(N^{2}LO\right)}\left(y\right)$}
eigenmodes back to the original $z$ coordinate via 
\begin{align}
\psi_{n}\left(z\right) & =\frac{z^{2}}{R^{3}}\sqrt{\frac{1}{R^{2}+z^{2}}}u_{n}\left(y\left(z\right)\right)
\end{align}
one ends up with uniform, global N$^{2}$LO solutions to the radial
field equation (\ref{eq:radeq}) which consist of three matching parts.
The first one is the near-horizon solution
\begin{equation}
\psi_{R,n}\left(z\right)\overset{z\rightarrow R}{=}c_{n}\frac{z^{2}}{R^{3}}\frac{e^{-\frac{1}{2}\sqrt{\frac{3}{\epsilon_{n}^{2}}-\frac{11}{2^{2}}}\frac{R^{2}-z^{2}}{z^{2}}}}{\left(R^{2}+z^{2}\right)^{1/2}}\,_{1}F_{1}\left(\frac{1}{2}-\frac{1}{4}\frac{\frac{2}{\epsilon_{n}^{2}}-3}{\sqrt{\frac{3}{\epsilon_{n}^{2}}-\frac{11}{2^{2}}}},1;\sqrt{\frac{3}{\epsilon_{n}^{2}}-\frac{11}{2^{2}}}\frac{R^{2}-z^{2}}{z^{2}}\right)\label{eq:psiR}
\end{equation}
(where $\,_{1}F_{1}$ is Kummer's confluent hypergeometric function
\cite{abr72}). In the intermediate region, the solution is
\begin{align}
\psi_{int,n}\left(z\right) & \overset{R\ll z\ll0}{=}c_{n}\sqrt{\frac{\epsilon_{n}}{\pi}}\frac{z^{3/2}}{R^{5/2}\left(R^{4}-z^{4}\right)^{1/4}}\times\nonumber \\
 & \times\cos\left\{ -\frac{2}{\epsilon_{n}}\left[\frac{z}{R}\,_{2}F_{1}\left(\frac{1}{4},\frac{1}{2},\frac{5}{4},\left(\frac{z}{R}\right)^{4}\right)-\frac{\pi^{3/2}}{2^{3/2}\Gamma\left(\frac{3}{4}\right)^{2}}\right]-\frac{\pi}{4}\right.\nonumber \\
 & -\frac{\epsilon_{n}}{16}\left.\left[\frac{15\frac{R}{z}-13\left(\frac{z}{R}\right)^{3}}{\sqrt{1-\left(\frac{z}{R}\right)^{4}}}+4\left(\frac{z}{R}\right)^{3}\,_{2}F_{1}\left(\frac{3}{4},\frac{1}{2},\frac{7}{4},\left(\frac{z}{R}\right)^{4}\right)-4\,_{2}F_{1}\left(\frac{3}{4},\frac{1}{2},\frac{7}{4},1\right)\right]\right\} .\label{eq:psiInt}
\end{align}
This is the only non-monotonic segment (for $n>1$) which therfore
contains the oscillatory part of the complete solution. Finally, near
the boundary the solution becomes
\begin{align}
\psi_{0,n}\left(z\right)\overset{z\rightarrow0}{=} & \left(-1\right)^{n-1}\frac{c_{n}}{2}\left(\frac{1}{\epsilon_{n}^{2}}+\frac{3}{2}\right)\frac{z^{4}}{R^{3}}\frac{e^{-\sqrt{\frac{3}{\epsilon_{n}^{2}}+\frac{9}{4}}\frac{z^{2}}{R^{2}-z^{2}}}}{\left(R^{2}+z^{2}\right)^{1/2}\left(R^{2}-z^{2}\right)}\nonumber \\
 & \times\,_{1}F_{1}\left(\frac{3}{2}-\frac{\frac{1}{\epsilon_{n}^{2}}+\frac{3}{2}}{2\sqrt{\frac{3}{\epsilon_{n}^{2}}+\frac{9}{4}}},3,2\sqrt{\frac{3}{\epsilon_{n}^{2}}+\frac{9}{4}}\frac{z^{2}}{R^{2}-z^{2}}\right).\label{eq:psi0}
\end{align}

The above solutions have the correct number of $n-1$ nodes, as expected
on general grounds and confirmed by the numerical solutions. (In contrast,
the NLO solutions have one additional, spurious node). As required
by normalizability, the modes exhibit sub-leading behavior near the
boundary, i.e.
\begin{equation}
\psi_{n}\left(z\right)\overset{z\rightarrow0}{\longrightarrow}c_{n}\left(-1\right)^{n-1}\frac{1}{2}\left(\frac{1}{\epsilon_{n}^{2}}+\frac{3}{2}\right)\frac{z^{4}}{R^{6}},\label{eq:nearboundbehav}
\end{equation}
and are regular at the horizon where they attain the finite values
\begin{equation}
\psi_{n}\left(R\right)=\frac{c_{n}}{\sqrt{2}R^{2}}.
\end{equation}
(Ortho-) normalizing the modes with respect to their Sturm-Liouville
inner product 
\begin{equation}
\int_{0}^{R}dz\sqrt{g}g^{zz}\psi_{m}\left(z\right)\psi_{n}\left(z\right)=\int_{0}^{R}dz\frac{R^{3}}{z^{3}}\psi_{m}\left(z\right)\psi_{n}\left(z\right)=\delta_{mn}\label{eq:orthonorm}
\end{equation}
fixes the overall coefficients $c_{n}$. The N$^{2}$LO approximations
to these intergrals can be derived from the $u_{n}\left(y\right)$
by exploiting the relation 
\begin{align}
\int_{0}^{R}dz\frac{R^{3}}{z^{3}}\left[\psi_{n}\left(z\right)\right]^{2} & =-\frac{1}{2R^{3}}\int_{-\infty}^{+\infty}dyV_{1}\left(y\right)\left[u_{n}\left(y\right)\right]^{2}
\end{align}
which simplifies the integration over the special functions and the
consistent two-scale power counting. The final result to N$^{2}$LO
accuracy, 
\begin{equation}
c_{n}^{2}\simeq\frac{\pi^{2}R^{3}}{\,_{2}F_{1}\left(\frac{1}{4},\frac{1}{2},\frac{5}{4},1\right)K\left(-1\right)}\left(n+\frac{1}{2}\right)=4\sqrt{2}\frac{\Gamma\left(\frac{3}{4}\right)^{3}}{\Gamma\left(\frac{1}{4}\right)}R^{3}\left(2n+1\right)\label{eq:c2n}
\end{equation}
(where $K\left(n\right)$ is the complete elliptic integral of the
first kind \cite{abr72}), is surprisingly simple and accurate (see
below). 

\begin{figure}
\begin{centering}
\includegraphics[height=7cm]{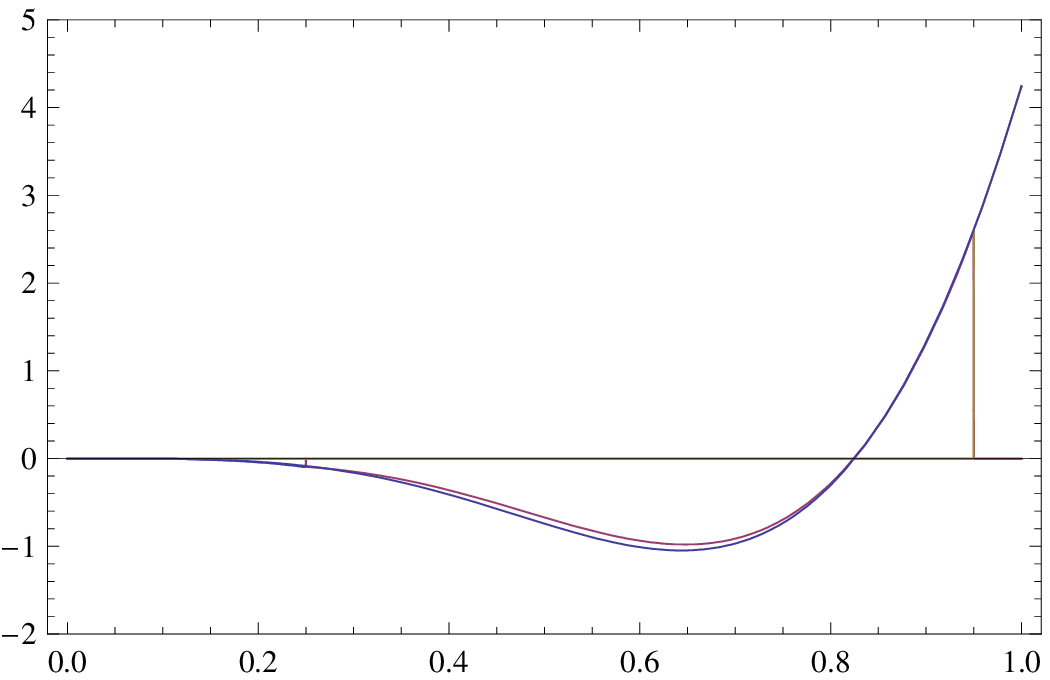} 
\par\end{centering}

\caption{The (unnormalized) numerical bulk-mode solution (lower curve) and
its N$^{2}$LO approximation (upper curve) for $n=2$. The two vertical
line segments indicate the matching points.}

\label{nEq2-WKBNum} 
\end{figure}

\begin{figure}
\begin{centering}
\includegraphics[height=7cm]{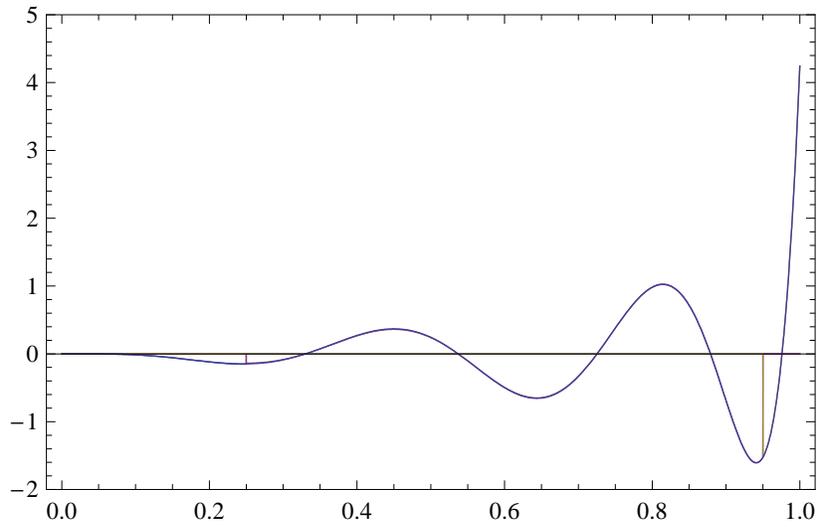} 
\par\end{centering}

\caption{Same as in Fig. \ref{nEq2-WKBNum}, but for $n=6$. (For $n\gtrsim4$
the numerical solution and the N$^{2}$LO approximation are indistinguishable
within plot resolution.) }

\label{nEq6-WKBNum} 
\end{figure}

By construction, the N$^{2}$LO solutions converge uniformly to the
exact solutions for small $\epsilon$, i.e. for large $n$. Their
accuracy turns out to be unexpectedly high already for the smallest
$n$, however. While the nodeless ground-state ($n=1$) mode is still
the least well approximated %
\footnote{In fact, it stays slightly discontinuous at the matching points. This
presents no relevant practical problems, however, since the ground-state
mode has the simplest $z$ dependence and can easily be improved (e.g.
variationally \cite{bro200}).%
}, the accuracy grows already to the sub-percent level for $n=2$,
as revealed by comparison with the numerical solution in Fig. \ref{nEq2-WKBNum}.
From $n\gtrsim4$ the N$^{2}$LO and numerical solutions are identical
within plot resolution (cf. Fig. \ref{nEq6-WKBNum}). Another rather
stringent test for the mode accuracy (and normalization) provides
the comparison of the N$^{2}$LO horizon values
\begin{equation}
\psi_{n}^{2}\left(R\right)=2\sqrt{2}\frac{\Gamma\left(\frac{3}{4}\right)^{3}}{\Gamma\left(\frac{1}{4}\right)}\frac{1}{R}\left(2n+1\right)\label{eq:psin(R)}
\end{equation}
with their numerical counterparts given in Tab. \ref{TableHorizonValue}.
The deviations start from about half a percent for $n=1$ and decrease
rapidly with increasing $n$. 
\begin{table}
\begin{tabular}{|c|c|c|c|c|c|c|c|c|c|c|}
\hline 
$n$ & $1$ & $2$ & $3$ & $4$ & $5$ & $6$ & $7$ & $8$ & $9$ & $10$\tabularnewline
\hline 
\hline 
$\psi_{n}^{\left(N^{2}LO\right)2}\left(R\right)$ & $4.30662$ & $7.17770$ & $10.0488$ & $12.9199$ & $15.7909$ & $18.6620$ & $21.5331$ & $24.4042$ & $27.2753$ & $30.1463$\tabularnewline
\hline 
$\psi_{n}^{\left(num\right)2}\left(R\right)$ & $4.28354$ & $7.17546$ & $10.0481$ & $12.9196$ & $15.7908$ & $18.6619$ & $21.5330$ & $24.4041$ & $27.2752$ & $30.1463$\tabularnewline
\hline 
\end{tabular}

\caption{Comparison of the N$^{2}$LO approximation (\ref{eq:psin(R)}) and
the numerical results for the value $\psi_{n}\left(R\right)$ of the
$n$-th eigenmode at the horizon (for $R=1$).}

\label{TableHorizonValue}
\end{table}

With the accurate string-mode approximations (\ref{eq:psiR}) - (\ref{eq:psi0})
whe have achieved our first and central objective. On their basis,
analytical glueball properties and amplitudes will be derived in the
following section.

\section{Analytical glueball properties}

The gauge-theory observables which are most directly related to the
normalizable eigenmodes are the glueball masses encoded in their eigenvalues.
From Eq. (\ref{eq:EVcond}) one finds, to N$^{2}$LO accuracy, the
mass spectrum 
\begin{equation}
m_{n}^{2}=\left[\frac{\pi\left(n+\frac{1}{2}\right)}{\,_{2}F_{1}\left(\frac{1}{4},\frac{1}{2},\frac{5}{4},1\right)R}\right]^{2}\simeq\frac{1}{2}n\left(n+1\right)m_{1}^{2},\qquad n\ge1\label{eq:mn2}
\end{equation}
which exhibits a finite gap %
\footnote{Note that the $n=1$ scalar glueball is the lowest-lying state of
the complete excitation spectrum \cite{bro00}.%
} \cite{min99} 
\begin{equation}
m_{1}^{2}=\frac{2^{4}}{\pi}\Gamma\left(\frac{3}{4}\right)^{4}\frac{1}{R^{2}}\simeq11.4843\frac{1}{R^{2}}\label{eq:m1}
\end{equation}
typical for color confinement %
\footnote{The quadratic large-$n$ relation $m_{n}^{2}\sim n^{2}$ is typical
for top-down gravity duals (whereas QCD generates linear trajectories).%
}. A fit to the first 50 numerical eigenvalues corroborates the $n$
dependence of the mass spectrum (\ref{eq:mn2}) and yields an almost
identical value $m_{1}^{\left(fit\right)2}=11.4844/R^{2}$ for the
mass gap. The quality of the N$^{2}$LO approximation is surprisingly
good again even for $n=1$, and excellent from about $n\gtrsim5$
(cf. Table \ref{TableMn2Fn2}). 

A second fundamental set of glueball observables are the pole residues
or ``decay constants'' $f_{n}$ which parametrize the matrix elements
\begin{align}
\langle0|O_{0^{++}}\left(x\right)|0_{n}^{++}\left(k\right)\rangle & =\langle0|O_{0^{++}}\left(0\right)|0_{n}^{++}\left(k\right)\rangle e^{-ikx}=f_{n}m_{n}^{2}e^{-ikx}.
\end{align}
In contrast to the ``local'' nature of the masses, the $f_{n}$
are of ``global'' origin in the sense that they depend on the overall
normalization of the modes,
\begin{equation}
f_{n}=\frac{1}{\kappa m_{n}^{2}}\lim_{\varepsilon\rightarrow0}\sqrt{g}\left(\varepsilon\right)g^{zz}\left(\varepsilon\right)\psi'_{n}\left(\varepsilon\right)=\frac{R^{3}}{\kappa m_{n}^{2}}\lim_{\varepsilon\rightarrow0}\frac{\psi'_{n}\left(\varepsilon\right)}{\varepsilon^{3}},
\end{equation}
(the prime indicates a radial derivative). Evaluating this formula
with the behavior (\ref{eq:nearboundbehav}) and normalization constants
(\ref{eq:c2n}) of the near-boundary modes (\ref{eq:psi0}) shows
that the N$^{2}$LO results can be cast into the form
\begin{equation}
f_{n}^{2}=\frac{1}{3}\left(2n+1\right)f_{1}^{2}\label{eq:fn2}
\end{equation}
with
\begin{align}
f_{1}^{2} & =3\sqrt{2}\frac{\Gamma\left(\frac{3}{4}\right)^{3}}{\Gamma\left(\frac{1}{4}\right)}\frac{R}{\kappa^{2}}\simeq2.153\frac{R}{\kappa^{2}}.\label{eq:f1}
\end{align}
Again, these results approximate their numerical counterparts very
well (cf. Tab. \ref{TableMn2Fn2}). A fit to the first 50 numerical
$f_{n}$ values confirms the overall $n$ dependence (\ref{eq:fn2})
and yields\textbf{ $f_{1}^{\left(fit\right)2}\simeq2.148R/\kappa^{2}$}
\footnote{In popular bottom-up models this behavior is not generally reproduced:
the $f_{n}$ grow e.g. similarly with $n$ in the ``hard-wall''
model while they become asymptotically $n$-independent in the ``soft
wall'' model \cite{for08}.%
}. The analytical form of the results reveals in addition a new relation
\begin{equation}
f_{n}^{2}=\frac{c_{n}^{2}}{2R^{2}\kappa^{2}}=\frac{R^{2}}{2\kappa^{2}}\psi_{n}^{2}\left(R\right)
\end{equation}
which corroborates that the residues contain information on the $\psi_{n}\left(z\right)$
over their whole domain $0\le z\le R$. 
\begin{table}
\begin{tabular}{|c|c|c|c|c|c|c|c|c|c|c|}
\hline 
$n$ & $1$ & $2$ & $3$ & $4$ & $5$ & $6$ & $7$ & $8$ & $9$ & $10$\tabularnewline
\hline 
\hline 
$m_{n}^{\left(N^{2}LO\right)2}$ & $11.48432$ & $34.45296$ & $68.90592$ & $114.84320$ & $172.26480$ & $241.17072$ & $321.56096$ & $413.43552$ & $516.79441$ & 631.63761\tabularnewline
\hline 
$m_{n}^{\left(num\right)2}$ & $11.58766$ & $34.52698$ & $68.97496$ & $114.91044$ & $172.33117$ & $241.23661$ & $321.62655$ & $413.50092$ & $516.85966$ & $631.70276$\tabularnewline
\hline 
$f_{n}^{\left(N^{2}LO\right)2}$ & $2.15331$ & $3.58885$$ $ & $5.02439$ & $6.45993$ & $7.89547$ & $9.33101$ & $10.7666$ & $12.2021$ & $13.6376$ & $15.0732$\tabularnewline
\hline 
$f_{1}^{\left(num\right)2}$ & $2.14172$ & $3.58746$ & $5.02404$ & $6.45978$ & $7.89540$ & $9.33097$ & $10.7665$ & $12.2021$ & $13.6376$ & $15.0732$\tabularnewline
\hline 
\end{tabular}

\caption{Comparison of the numerical and N$^{2}$LO results for $m_{n}^{2}$
and $f_{n}^{2}$. }

\label{TableMn2Fn2}
\end{table}

For related reasons, the pole residues also contain information on
the size of the associated gauge-theory bound states. The increasing
support of the dual modes (\ref{eq:psiR}) - (\ref{eq:psi0}) near
the horizon in the fifth dimension may e.g. indicate that the glueball
sizes remain bounded for large excitation number $n$. A more explicit
relation between $f_{n}$ and the glueball sizes arises from the (properly
renormalized) coincidence limit of the gauge-invariant Bethe-Salpeter
amplitudes 
\begin{equation}
\chi_{n}\left(x\right)=\left\langle 0\left|tr\left\{ F_{\mu\nu}\left(-\frac{x}{2}\right)U\left(-\frac{x}{2},\frac{x}{2}\right)F^{\mu\nu}\left(\frac{x}{2}\right)\right\} \right|0_{n}^{++}\right\rangle 
\end{equation}
(where $U$ is the adjoint color parallel transporter). It shows that
the $f_{n}m_{n}^{2}\sim\lim_{\left|x\right|\rightarrow0}\chi_{n}\left(x\right)$
play the role of the $n$-th glueball's ``wave function at the origin''
\footnote{The precise interpretation of the Bethe-Salpeter ``density'' is
not straightforward, however. A proper size definition should therefore
be based on form factors describing conserved charge distributions.%
}.

Finally, we demonstrate how the GWKB eigenmode approximations (\ref{eq:psiR})
- (\ref{eq:psi0}) can be used to obtain accurate analytical approximations
for more complex gauge-theory amplitudes. An illustrative example
is provided by the two-point correlation function of the interpolator
(\ref{eq:GbOp}). Its generating functional (\ref{eq:gfun}) is determined
by the (Gaussian) on-shell action which reduces Eq. (\ref{eq:DilatonAction})
to the surface term 
\begin{equation}
S_{\partial M}^{\left(onsh\right)}\left[\varphi\right]=-\frac{1}{2\kappa^{2}}\int_{\partial M}d^{4}x\left[\sqrt{g}g^{zz}\varphi\left(x,z\right)\partial_{z}\varphi\left(x,z\right)\right]_{z\rightarrow0}\label{eq:OnShellAction1}
\end{equation}
where $\varphi\left(x,z\right)$ is a solution of the field equation
(\ref{eq:LBeq}) (which causes the bulk action to vanish). The (boundary)
Fourier-transformed solutions corresponding to a finite boundary source
$\hat{\varphi}^{\left(s\right)}\left(q\right)$ can be written as
\begin{equation}
\hat{\varphi}\left(q,z\right)=\hat{K}\left(q,z\right)\hat{\varphi}^{\left(s\right)}\left(q\right)\label{eq:gensol}
\end{equation}
where $\hat{K}\left(q,z\right)$, subject to the UV boundary condition
$\hat{K}\left(q;\varepsilon\rightarrow0\right)=1$, is the bulk-to-boundary
propagator \cite{wit298}. Inserting the solution (\ref{eq:gensol})
into Eq. (\ref{eq:OnShellAction1}) casts the on-shell action into
the form
\begin{align}
S_{\partial M}^{\left(onsh\right)}\left[\varphi_{s}\right] & =\frac{1}{2}\int\frac{d^{d}q}{\left(2\pi\right)^{d}}\hat{\varphi}_{s}\left(-q\right)\Pi\left(q\right)\hat{\varphi}_{s}\left(q\right)
\end{align}
where the momentum-space correlator is given by 
\begin{align}
\Pi\left(q\right) & =i\int d^{3}xe^{iq\left(x-y\right)}\left\langle T\mathcal{O}_{0^{++}}\left(x\right)\mathcal{O}_{0^{++}}\left(y\right)\right\rangle =\frac{1}{\kappa^{2}}\left[-\sqrt{g}g^{zz}\hat{K}\left(q,z\right)\partial_{z}\hat{K}\left(-q,z\right)\right]_{z=\varepsilon\rightarrow0}.\label{eq:Pi(q)}
\end{align}

We now expand $\hat{K}$ into the complete set of normalizable bulk
eigenmodes $\psi_{n}$, which results in the spectral representation
\begin{equation}
\hat{K}\left(q,z\right)=\kappa\sum_{n}\frac{f_{n}m_{n}^{2}}{q^{2}+m_{n}^{2}}\psi_{n}\left(z\right).\label{eq:ksprep}
\end{equation}
(The coefficents are obtained by noting that both $\hat{K}\left(q,z\right)$
and the $\psi_{n}\left(z\right)$ solve the radial field equation
(\ref{eq:radeq0}) and that the $\psi_{n}\left(z\right)$ are orthonormal
under the inner product (\ref{eq:orthonorm}).) Plugging Eq. (\ref{eq:ksprep})
into Eq. (\ref{eq:Pi(q)}) then provides the (formal) spectral representation
\begin{equation}
\Pi\left(q\right)=\sum_{n}\frac{f_{n}^{2}m_{n}^{4}}{q^{2}+m_{n}^{2}}=:\Pi^{\left(3\right)}\left(q^{2}\right)+P^{\left(3\right)}\left(q^{2}\right)
\end{equation}
for the 2-point correlation function. It contains a divergent subtraction
polynomial $P^{\left(3\right)}$ which can be extracted as 
\begin{equation}
P^{\left(3\right)}\left(q^{2}\right)=\left.\Pi\left(q^{2}\right)-q^{2}\frac{d\Pi\left(q^{2}\right)}{dq^{2}}+\frac{1}{2}q^{4}\frac{d^{2}\Pi\left(q^{2}\right)}{d\left(q^{2}\right)^{2}}\right|_{q^{2}=0}=\sum_{n=1}^{\infty}f_{n}^{2}m_{n}^{2}\left(1-\frac{q^{2}}{m_{n}^{2}}+\frac{q^{4}}{m_{n}^{4}}\right).
\end{equation}
The remainder, i.e. the thrice subtracted correlator $\Pi^{\left(3\right)}$,
is then given in terms of the finite spectral sum 
\begin{align}
\Pi^{\left(3\right)}\left(q^{2}\right) & =-\sum_{n=1}^{\infty}\frac{q^{6}f_{n}^{2}}{m_{n}^{2}\left(q^{2}+m_{n}^{2}\right)}.\label{eq:Pi3}
\end{align}
Now it is important to observe that the N$^{2}$LO expressions (\ref{eq:mn2}),
(\ref{eq:m1}) for the masses and (\ref{eq:fn2}), (\ref{eq:f1})
for the pole residues allow Eq. (\ref{eq:Pi3}) to be summed analytically.
The result is 
\begin{align}
\Pi^{\left(3\right)}\left(q^{2}\right) & =-\frac{2f_{1}^{2}}{3m_{1}^{2}}q^{4}\left[2\gamma_{E}-1+\frac{m_{1}^{2}}{2q^{2}}+\psi\left(\frac{1}{2}+\frac{\sqrt{1-8q^{2}/m_{1}^{2}}}{2}\right)+\psi\left(\frac{1}{2}-\frac{\sqrt{1-8q^{2}/m_{1}^{2}}}{2}\right)\right]\label{eq:Pi3-2}
\end{align}
(where $\psi\left(x\right)$ is the digamma function \cite{abr72}).
Equation (\ref{eq:Pi3-2}) seems to be the first explicit result for
a correlation function in the $D3$-brane dual for YM$_{3}$. It provides
a uniformly accurate approximation to the exact correlator (evaluated
numerically) and a generalizable paradigm \cite{for16} for obtaining
analytical gauge-theory amplitudes from top-down gravity duals %
\footnote{The exact summation over all excitations requires analytical expressions
for all poles and residues with their explicit $n$ dependence. A
numerical treatment, trucated to a finite set of low-lying modes,
is generally not sufficient.%
}.

The coefficient in front of Eq. (\ref{eq:Pi3-2}) is determined by
the GWKB results (\ref{eq:m1}) and (\ref{eq:f1}) for the lightest
glueball as 
\begin{equation}
\frac{f_{1}^{2}}{m_{1}^{2}}=\frac{3}{2^{4}}\frac{R^{3}}{\kappa^{2}}.
\end{equation}
The Fourier-transformed correlator $\Pi\left(x\right)$ shows the
expected exponential decay $\Pi\left(x\right)\overset{\left|x\right|\rightarrow\infty}{\longrightarrow}\sim\exp\left(-m_{1}\left|x\right|\right)$
generated by the mass gap (\ref{eq:m1}). The large $q^{2}/m_{1}^{2}\gg1$
expansion 
\begin{align}
\Pi^{\left(3\right)}\left(q^{2}\right) & =-\frac{1}{2^{3}}\frac{R^{3}}{\kappa^{2}}q^{4}\left[2\gamma_{E}-1+\ln2+\ln\frac{q^{2}}{m_{1}^{2}}+\frac{1}{3}\frac{m_{1}^{2}}{q^{2}}-\frac{1}{60}\frac{m_{1}^{4}}{q^{4}}-\frac{25771}{2^{8}3^{2}5\times7}\frac{m_{1}^{6}}{q^{6}}+...\right]\label{eq:largeQ}
\end{align}
provides a rather good approximation to the correlator (\ref{eq:Pi3-2})
already for $q^{2}\gtrsim3m_{1}^{2}$ and bears some resemblance with
an operator product expansion (OPE) \cite{for00} (although the gravity
dual remains strongly coupled in the UV). Besides the conformal logarithm
(with prefactor $\sim q^{4}$ as in YM$_{4}$) and ``condensate-induced''
power corrections, the expansion (\ref{eq:largeQ}) also contains
a term $\sim1/q^{2}$ which cannot appear in the OPE. A more thorough
analysis and discussion of glueball correlators in top-down duals,
obtained by the above methods, will be given elsewhere \cite{for16}.

\section{Summary and conclusions}

The main purpose of this letter was to outline a controlled approximation
scheme for the analytical treatment of string-based gauge/gravity
duals. Our approach is based on uniform high-accuracy approximations
for normalizable string modes which are dual to gauge-theory states.
The approximate mode solutions are obtained from a generalized WKB
expansion driven to at least next-to-next to leading order. Analytical
results for gauge-theory properties and amplitudes are then calculated
using spectral methods.

We have demonstrated the potential of this approach in one of the
simplest top-down models for strong-interaction physics, i.e. in the
non-extremal, compactified $D3$-brane string dual for Yang-Mills
theory in three dimensions (with gauge group $SU\left(N\right)$ at
large $N$ and at strong 't Hooft coupling $\lambda=g_{YM}^{2}N$).
We derived controlled, global approximations for the dilaton modes
dual to scalar glueballs and obtained physically transparent analytical
approximations for the associated gauge-theory amplitudes, including
the two-point glueball correlation function.

New results and insights from our analytical treatment comprise (i)
the systematics of the mode behavior with increasing radial excitation
number, (ii) new relations between glueball properties and global
aspects of their dual modes, (iii) highly accurate analytical expressions
for the pole residues (or ``decay constants''), (iv) insights into
the size systematics of the scalar glueball excitations and (v) the
first expression for the scalar glueball correlator for which we obtain
a uniformly accurate analytical approximation. The latter demonstrates
the potential of deriving controlled analytical approximations to
``hadronic'' amplitudes by exact summation of their spectral representations.

We end by noting several general benefits of our approach. The first
is the physical transparency of many results which reveals the systematics
of contributions from towers of excited states. The latter are efficiently
condensed into comprehensive and often instructive analytical expressions.
This represents a clear advantage over numerical solutions. A second
general benefit is that the underlying set of approximate string-mode
solutions contains all glueball-related information needed for the
calculation of other\textbf{ }glueball amplitudes, including e.g.
form factors and scattering amplitudes. A third strength of our method,
finally, is the universality with which it can be adapted to rather
different and more complex top-down gravity duals, including e.g.
those which do not approach AdS spaces at their UV boundary. 
\begin{acknowledgments}
The author would like to thank the CERN Theory Group for hospitality
and financial support during the initial phase of this work. 
\end{acknowledgments}
\newpage{}

\newpage{}

\end{document}